\documentstyle[aps,preprint,psfig]{revtex}
\begin{document}
 \title{Measuring the Topology of the Universe}
\author{Neil J. Cornish, \\
Department of Applied Mathematics and Theoretical Physics, Cambridge
University, UK \\ David N. Spergel, \\
Princeton University Observatory, Princeton, NJ 08544 \\
\& Glenn D. Starkman 
\\
Case Western Reserve University, Cleveland, OH}
\maketitle
\centerline{{\it Classification:} Physical Sciences: Astronomy}
\section{Abstract}

Observations of microwave background fluctuations can yield information not
only about the geometry of the universe, but potentially about the topology
of the universe. If the universe is negatively curved, then the
characteristic scale for the topology of the universe is the curvature
radius. Thus, if we are seeing the effects of the geometry of the universe,
we can hope to soon see signatures of the topology of the universe. The
cleanest signature of the topology of the universe is written on the
microwave sky: there should be thousands of pairs of matched circles. These
circles can be used to determine the precise topology and volume of the
universe. Since we see hundreds of slices through the fundamental domain of
the universe, we can use the microwave observations to reconstruct the
initial conditions of the entire universe on the scale of a few Megaparsecs.

\section{Introduction}

There is growing evidence that we live in a negatively curved universe. A
number of independent arguments suggest that the matter density is
significantly less than the critical density: comparisions of density
fluctuations and velocity fields (Willick \& Strauss 1995; Davis, Nusser \& Willick 1996;
Riess et al. 1997); determination of mass-to-light ratios
in clusters (Bahcall, Lubin \& Dorman 1995);
measurements of the baryon to dark matter ratio in
clusters (S. White et al. 1993; Lubin et al. 1996; D. White, Jones \& Forman 1997); the presence of more large scale structure than expected in
flat models (Da Costa et al. 1994; Lin et al. 1996);
the need to reconcile the value of the Hubble constant
with globular cluster ages (Spergel, Bolte \& Freedman 1997);
the existence of large
numbers of clusters at moderate redshift (Carlberg et al. 1997; Bahcall, Fan \& Cen 1997). A number of
groups have shown that CDM models with $\Omega _0\simeq 0.3-0.4$ are
compatible with microwave background measurements and large scale structure
(Kamionkowski \& Spergel 1994; Ratra et al. 1997). While we cannot rule out
the possibility that there is a cosmological constant (e.g., Steinhardt \&
Ostriker 1995) that makes the universe flat, recent high redshift supernova
observations (Perlmutter et al. 1997), as well as gravitational lens statistics
(Turner 1990;
Falco, Kochanek \& Munoz 1997), suggest that this term is small.

There has been growing interest in the possibility that the universe is not
only negatively curved, but compact (Gott 1980; Fagundes 1983; Cornish, Spergel 
\& Starkman 1996a, 1996b; Bond, Pogosyan \& Souradeep 1997;
Levin et al. 1997). Our
interest in the topology of the universe was stimulated by the possibility
that it may be detectable and by the philosophical attractions of a finite
universe (Cornish, Spergel \& Starkman 1996a). This talk reviews some basic concepts in topology and then
turns to the possibility of detecting the observational signature of a
finite universe.

\section{Topology: A Quick Primer}

Physicists assume that the universe can be described as a manifold.
Mathematician characterize manifolds in terms of their geometry and
topology. Geometry is a local quantity that measures the intrinsic curvature
of a surface. General relativity relates the mass distribution of the
universe to its geometry and of course, the geometry of the universe
determines the dynamics of the mass. Topology is a global quantity that
characterizes the shape of space (see, e.g., Weeks 1985 for a general
introduction). General relativity does not at all constrain the topology of
the universe.

The relationship between topology and geometry is most familiar in flat
space. Cosmologists often consider flat infinite universes: a model that
mathematicians denote as $R^3$, which symbolizes a space that is a the
product of the three orthogonal real lines. A familiar cosmological model
that has the same geometry as $R^3,$ but a different topology is the three
torus or what mathematician call $T^3.$ Most cosmological simulations are
run on a three torus: if a particle tries to leave the computational
cube through one side it emerges on the opposite side.

There are several different ways of thinking about the topology of a
manifold. It is simplest to begin by considering topology in flat two
dimensional space. One example of a flat topology is a
square with identified sides. This square is the \textit{fundamental domain}
of the topology. Another way of thinking about this topology is to
glue the sides
together to create a donut, a two dimensional surface embedded in a three
dimensional space. Yet another way of thinking about this topology is
to tile an
infinite plane with identical copies of the same fundamental domain.

Cosmologists generally consider three possible geometries for the universe:
a positively curved universe, a flat universe, and a negatively curved
universe. In standard parlance, closed, critical and open universes. This
later nomenclature is misleading: negatively curved universes can be either
infinite in spatial extent or compact. Both models have the same dynamics
and expand forever: dynamics is determined by geometry. The three-sphere is
a space of constant positive curvature, and the pseudosphere is a hyperbolic
space with constant negative curvature.

From a topological point of view, negatively curved (\textit{hyperbolic})
universes are ``generic'': most three dimensional manifolds can be viewed as
homogenous negatively curved and compact (Thurston's geometrization
conjecture [Thurston 1978]). Cornish, Gibbons and Weeks (1997) have
recently shown that in the Hartle-Hawking approach to quantum cosmology, the
smallest volume manifolds are favored. These smallest volume manifolds are
the simplest (least complex) compact models.

In negatively curved manifolds, the characteristic length is the curvature
scale. Thus, throughout the rest of this talk, we will use it as our unit
of length. There are an infinite number of distinct compact hyperbolic
topologies. There is a fundamental group, usually denoted $\Gamma $ ,
associated with each of these topologies. Each of these topologies also has
a specific volume (measured in curvature units). We strongly recommend the
publicly available SNAPPEA program (http://www.geom.umn.edu/) for anyone
interested in developing a more intuitive feel for the rich structure of
compact hyperbolic topology. 

One of the intriguing mathematical properties of compact hyperbolic
manifolds is that geodesic flows on these manifolds are maximally
chaotic and mixing. Because of this property they are extensively studied by
physicists and mathematicians interested in quantum chaos(Balazs \& Voros 1986;
Gutzweiler 1985). We
have speculated that quantum chaos plays an important role in the
homogenization of the early universe (Cornish, Spergel \& Starkman 1996a).

\section{Observing the Effects of Topology}

How would we know if we are living in a compact hyperbolic universe? Local
observations can only measure the geometry of the universe. We need to look
out beyond our own fundamental domain to begin to see the effects of
topology. Most cosmologists trying to look for observational signature of
topology have attempted to find replicas of our own Galaxy or other familar
objects such as rich clusters (Gott 1980; Lehoucq,
Lachieze-Rey \& Luminet 1996;
Roukema \& Edge 1997). In a typical small volume
compact hyperbolic universe, the expectation value for the length of the
shortest closed geodesic in generic small volume manifolds (volumes less
than 10 say) is between 0.5 and 1 (Thurston, private communication). That
is, if we randomly select a point in such a manifold, this is the most
probable value for the conformal distance to our nearest copy.
In a matter dominated universe, the redshift of an object is related to the
conformal distance by
\begin{equation}
1+z=\frac{2(\Omega^{-1}_0-1)}{\cosh (\eta _0-\eta )-1} 
\end{equation}
where $\eta $ is the conformal lookback time, and $
\eta _0={\rm arccosh}(2/\Omega _0 -1) $
is the present conformal time. For an $\Omega _0=0.3$ universe, $\eta _0=2.42
$ and the nearest image of ourselves is likely to lie between a redshift of
0.9 and $2.9.$ Similarly, if $\Omega _0=0.4$, we find $\eta _0=2.06$ and
the nearest copy will typically lie between a redshift of $1.0$ and $3.8$.
These numbers indicate that it will be very difficult to constrain the
topology of a hyperbolic universe using direct searches for ghost images of
any astrophysical objects. This difficulty is compounded by the evolution of
astrophysical objects on much shorter timescales.

Fortunately, observations of the microwave background are potentially a very
powerful tool for probing the topology of the universe. MAP\ and Planck will
measure millions of independent points on the surface of last scatter. If we
live in a universe where the distance to our nearest copy is less than the
diameter of the last scattering surface, then we will see the same position
on the surface of last scatter at multiple points on the microwave sky
(Cornish, Spergel \& Starkman 1996a,b).
Since the surface of last scatter is a sphere, it will intersect itself
along circles. This will lead to pairs of matched circles across the sky.
Note that the temperature is not constant along these circles, but rather
there are pairs of points along each circle with identical temperatures.

There are several effects that make this signature very difficult to detect
in the COBE\ data. If the universe is negatively curved, then most of the
large-scale fluctuations are not due to physics at the surface of last
scatter, but rather due to the decay of potential fluctuations along the
line of sight (Kamionkowski \&\ Spergel 1994). These fluctuations are mostly
generated at $z<2,$ generally within our fundamental domain. Thus, large
angular scale measurements are not sensitive to topology. Even without this
effect, we will likely need higher resolution maps with higher
signal-to-noise to definitively detect the matched circles. The combination
of high signal-to-noise and large numbers of independent pixels along each
circle significantly reduces the chances of false detections.

Can we detect topology with the MAP data? The number of matched circle pairs
is equal to the number of copies of the fundamental cell that can fit inside
a ball of proper radius $\chi =2\eta _0$ (we are approximating the radius of
the LSS by the radius of the particle horizon). A fair estimate of the
number of cells, $N_{c\text{,}}$ needed to tile this hyperbolic ball is
given by the volume ratio, 
\begin{equation}
N_c=\frac{\pi \left[ \sinh (4\eta _0)-4\eta _0\right] }{\text{Vol}(\Sigma )}
\end{equation}
where Vol$(\Sigma )$ is the volume of the fundamental domain in curvature
units. In a universe with $\Omega _0=0.3,$ this is a whopping ratio of $%
N_c=25000/$Vol$(\Sigma )$. Considering that there are an infinite number of
hyperbolic three manifolds with volumes less that $3$, we have no shortage
of model universes that can be seen by this method. Even if $\Omega _0$ was
as high as $0.95$, we would still have an infinite number of manifolds to
chose from that would produce at least some matched circle pairs.

How big are these matched circles? Lets consider an infinite hyperbolic
universe tiled with copies of our fundamental domain. We can draw the last
scattering sphere centered not only around our own location, but also around
all of our images. The circles arise at the intersections of these spheres.
The angular radius, $\theta ,$ of the circles is set by the radius of the
last scattering sphere ($\sim \eta _0)$, and the conformal distance, $%
\xi ,$ to the appropriate image of the MAP satellite: 
\begin{equation}
\theta =\cos ^{-1}\left( \frac{\cosh \xi -1}{\tanh \eta _0\sinh \xi }\right) 
\end{equation}
If $\Omega _0=0.3,$ there will be over $4000/$Vol($\Sigma )$ circles of
radius between 10$^o$ and 15$^o$ degrees. At its highest frequency, MAP will
have a resolution of 0.21$^o$, thus it will measure over 300 independent
points along each of these circles. Even at a resolution of 0.5$^o,$ where
we can use the three highest frequency maps, there are still over $125$
independent points along each circle. We are currently simulating the
analysis of MAP data using synthetically generating skies: our current
results are promising, it appears that MAP\ has sufficient signal-to-noise
and resolution to detect topology in a $\Omega =0.3$ universe even if the
volume of the fundamental domain of several tens of curvature volumes.

Once we have detected the circles, J. Weeks showed us that we can use them
to determine the topology of the universe. Each pair of circles lie on
matched faces of the fundamental cell, thus, each pair of circles gives us
an element of the fundamental group $\Gamma $. From the list of elements, we
can construct the generators of the group. Most small volume universe have
only two or three generators. Since we expect to have many hundreds if not
thousands of circle pairs, constructing the generators from the elements is
a highly overdetermined problem. Thus, producing a consistent solution is
an extremely powerful check that will help demonstrate that the circles are not
random events or artifacts of some pernicious instrumental effect. The
Mostow-Prasad rigidity theorem implies that a given topology has a fixed
volume measured in curvature units, thus, once we know the topology and the
angular size of the circles, we have an independent topological determination
of the radius of last scattering surface in units of the curvature radius.
This yields an independent measurement of $\Omega $ that does not rely on
any assumptions about the nature of the primordial fluctuations.

After determining the topology of the universe, we can then proceed onwards
and reconstruct the temperature of the photons and the velocity of the baryon
photon fluid throughout the entire fundamental domain from the microwave
sky. The surface area of the microwave sky is $4\pi \sinh ^2\eta _0,$ thus,
in an $\Omega _0=0.3$ universe, we see $\sim 400/$Vol$(\Sigma )$ slices
through the fundamental domain. Hence, the characteristic distance between
slices is only a few co-moving Megaparsecs. We should be able to reconstruct
not only the statistical properties of the initial fluctuations, but the
actual amplitude and phases of the initial fluctuations on the scale of
galaxy clustering. These initial fluctuations can, of course, be integrated
forward and compared to the local observed universe. At the end of the
analysis, we, in principle, should be able to identify the cold spot that
eventually collapsed to form the Virgo supercluster and other familiar parts
of our local universe. Cosmologists are used to thinking of looking out at the
universe and measuring the prehistory of other regions of the universe. If
we are fortunate enough to live in a compact hyperbolic universe, we can
look out and see our own beginnings.

\section{Acknowledgments}

We would like to thank Jeff Weeks and Bill Thurston for deepening our
understanding of topology and for helpful comments. DNS\ is supported by the
MAP/MIDEX program. GDS is supported through the NSF\ CAREER grant.
NJC was supported by PPARC grant GR/L21488.

\end{document}